\documentclass[aps,preview,twocolumn,showpacs]{revtex4}
\usepackage{stmaryrd}
\usepackage{amsfonts}
\usepackage{mathrsfs}
\usepackage{amsmath}
\usepackage{amscd}
\usepackage{graphicx}
\usepackage{booktabs}
\usepackage{leftidx}
\begin{document}
\title{Protected quantum state transfer in decoherence-free subspaces}
\author{Wei Qin$^1$}\protect\thanks{Corresponding author:  qinwei09@tsinghua.org.cn}
\author{Chuan Wang$^2$}
\author{Xiangdong Zhang$^1$}
\address{$^1$ School of Physics, Beijing Institute of Technology, Beijing 100081, China\\
$^2$ State Key Laboratory of Information photonics and Optical
Communications, Beijing University of Posts and Telecommunications, Beijing 100876,
China}

\begin{abstract}
We propose and analyse a robust quantum state transfer protocol by the use of a combination of coherent quantum coupling and decoherence-free subspaces in a coupled quantum spin chain. Under decoherence, an arbitrary unknown quantum state embedded in a decoherence-free subspace can be perfectly transferred through a noisy channel being weakly coupled to two end registers. The method protects quantum information from both the channel noise and the environmental decoherence. A special case of utilizing two physical qubits to encode one logical qubit is considered as an example and the robustness is confirmed by numerical performances.
\end{abstract}
\pacs{03.67.Lx, 03.67.Pp, 75.10.Pq}
\maketitle
\section{introduction}
\label{se:section1}
The realization of reliable quantum state transfer (QST) between two distant quantum registers forms essentially a fundamental building block for any scalable quantum information processing. To this end, the registers are mediated by quantum channels, which have been explored in the context of various quantum systems \cite{superconductor,ion,semiconductor,photon1,photon2}. Among them, coupled quantum spin chains have attracted much attention as interesting alternatives to either direct qubit interactions or interface between stationary and flying qubits for short distance quantum communication in recent decades \cite{spin1,spin2,spin3,spin4,spin5,wspin1,wspin2,wspin3,wspin4,wspin5,wspin6}. In this case, noise in the intermediate channel is a major practical hurdle in achieving a perfect QST. One straightforward approach to avoiding this noise is to initialize the channel in a pure state, e.g., a ferromagnetic state with spins aligning in a parallel way \cite{spin1,spin2,spin3,spin4,spin5}. However, it commonly relies upon high control requirements and is not easily granted in experiments. With accessibility to experimental implementations, another novel solution is to employ a coherent quantum coupling mechanism \cite{weakspin1}, which can enable a high efficient QST through a randomized channel in a maximally mixed state \cite{weakspin2}, corresponding to the infinite temperature limit. So far, this mechanism has been widely used in various QST protocols \cite{weakspin3,weakspin4,weakspin5,weakspin6,weakspin7}.

Besides noise in the intermediate channel, environmental decoherence is also a major practical hurdle in achieving a perfect QST \cite{decospin1,decospin2}.
It is caused by the inevitable interaction between an open system and its surrounding environment, and can destroy the desired coherence of the system. Many strategies addressing this challenge have been focused on using either quantum error corrections \cite{qer1,qer2,qer3}, dynamical decoupling \cite{ddc1,ddc2} or decoherence-free subspaces (DFSs) \cite{dfs1,dfs2}. In these approaches, the proposal of DFSs is a passive candidate to encode quantum information in a noiseless subspace of the overall Hilbert space, over which the states undergo unitary evolution. The basic condition for the feasibility of DFSs is that the system is subject to a symmetric coupling to its environment. Until now, DFSs have been extensively studied both in theoretical and experimental frames \cite{qdfs1,qdfs2,qdfs3,qdfs4,qdfs5}.

The goal of this paper is to provide a theoretical basis that quantum information could be protected from both the channel noise and the environmental decoherence. In this paper, we develop a hybrid approach to combine the coherent quantum coupling and the DFSs in an XX coupling spin chain. Specifically, two multi-qubit registers interact weakly with a noisy channel, and quantum information is encoded in the DFSs of the registers. Under decoherence characterized by a dephasing model for an open system collectively coupled to its surrounding environment via quantum nondemolition interaction, time evolution for a specific period results in a high fidelity swap operation. In a special case where two physical qubits are harnessed to encode one logical qubit, numerical simulations confirm that the protected QST is achievable. Observing these is of both practical and fundamental importance in the quest for experimental implementations of scalable quantum devices.

The paper is organized as follows. In Sec. \ref{se:section2}, we study the coherent coupling and QST between two registers. Sec. \ref{se:section3} presents a perfect QST protocol by combining the coherent coupling and the DFS methods. In Sec. \ref{se:section4}, a special case of two-qubit registers is considered as an example to confirm the robust QST. And the last section is our summary.

\section{The Model of quantum coherent coupling}
\label{se:section2}
\begin{figure}[!ht]
\begin{center}
\includegraphics[width=8.5cm,angle=0]{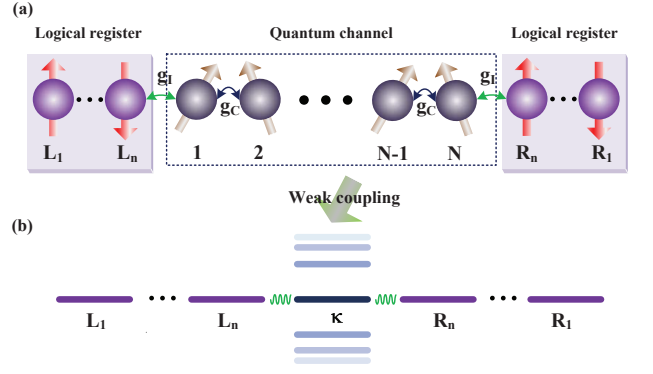}
\caption{(Color online) (a) Two distant quantum registers, each of which consists of $n$ interacting physical qubits, are mediated by an $N$-spin chain serving as a quantum channel. (b) Under the weak-coupling approximation, $g_{I}/g_{C}\ll1$, the two registers are resonantly coupled to a zero-energy collective mode of the intermediate channel. By encoding quantum information in the DFSs, we achieve a protected quantum state transfer against both the channel noise and the environmental decoherence.}\label{f1}
\end{center}
\end{figure}

Consider a spin chain of $N$ qubits working as a quantum channel to connect two additional spin chains as quantum registers termed $L$ and $R$, each of which consists of $n$ interacting spins denoted by $L_{1},\cdots,L_{n}$ and $R_{1},\cdots,R_{n}$, respectively, as depicted in Fig. \ref{f1}(a). The Hamiltonian of the system includes three parts, $H_{0}=H_{\text{ch}}+H_{\text{reg}}+H_{\text{reg-ch}}$. $H_{\text{ch}}$ is an XX coupling of the quantum channel
\begin{equation}
H_{\text{ch}}=g_{C}\sum_{i=1}^{N-1}\left(\sigma_{i}^{+}\sigma_{i+1}^{-}+\sigma_{i}^{-}\sigma_{i+1}^{+}\right),
\end{equation}
where $\sigma^{\pm}_{i}=\left(\sigma^{x}_{i} \pm i\sigma^{y}_{i}\right)/2$, and $\boldsymbol{\sigma}_{i}=\left\{\sigma^{x}_{i},\sigma^y_{i},\sigma^z_{i}\right\}$ are the Pauli spin matrices for the $i$th spin. $H_{\text{reg}}$ characterizes the Hamiltonian for the two registers
\begin{equation}
H_{\text{reg}}=\sum_{u=1}^{n-1}g_{u}\left(\sigma_{L_{u}}^{+}\sigma_{L_{u+1}}^{-}+\sigma_{R_{u}}^{+}\sigma_{R_{u+1}}^{-}\right)+\text{H.c.}
\end{equation}
where $g_{u}$ is the coupling strength between two nearest-neighboring qubits, and $\text{H.c.}$ stands for the Hermitian conjugate. The interaction between the end registers and the intermediate channel is
\begin{equation}
H_{\text{reg-ch}}=g_{I}\left(\sigma_{L_{n}}^+\sigma_{1}^{-}+\sigma_{R_{n}}^{+}\sigma_{N}^{-}\right)+\text{H.c.},
\end{equation}
with a register-channel coupling strength $g_{I}$. Compactly, the Hamiltonian $H_{0}$ can be rewritten as $H_{0}=\sum_{i,j} \Omega_{ij}\sigma_{i}^{+}\sigma_{j}^{-}$. Here, $\Omega$ is expressed as an $\left(N+4\right)\times\left(N+4\right)$ coupling matrix, and the summations run from $L_{1}$ to $R_{1}$.  Owing to the relation $\left[\sum_{i}\sigma_{i}^{z},H_{0}\right]=0$, $H_{0}$ conserves the total $z$ component of spins. Following the Jordan-Wigner transformation \cite{JWT},
\begin{equation}
\sigma_{j}^{+}=a_{j}^{\dag}\text{exp}\left(i\pi\sum_{l<j}a_{l}^{\dag}a_{l}\right),\nonumber
\end{equation}
yields a fermionized tight-binding Hamiltonian
\begin{equation}\label{eq:fermH}
H_{0}=\sum_{i,j}\Omega_{ij}a_{i}^{\dag}a_{j}, \nonumber
\end{equation}
where $a_{i}^{\dag}$ and $a_{i}$ are fermionic creation and annihilation operators, respectively, and they satisfy fermionic anticommutation relations, i.e., $\left\{a_{i},a_{j}\right\}=\left\{a_{i}^{\dag},a_{j}^{\dag}\right\}=0$ and $\left\{a_{i},a_{j}^{\dag}\right\}=\delta_{ij}$. In the Heisenberg picture, time evolution of an operator is driven by the Heisenberg equation of motion, $\dot{a}_{i}\left(t\right)=i\left[H_{0},a_{i}\left(t\right)\right]$, and by applying this rule, we evaluate
\begin{equation}
a_{i}\left(t\right)=\sum_{j}\Delta_{ij}a_{j},
\end{equation}
with $\Delta=e^{-i\Omega t}$ being a unitary evolution matrix.

In the following, the register-channel coupling is supposed to be much smaller than intrachannel coupling, $g_{I}/g_{C}\ll1$, such that $H_{\text{ch}}$ is perturbed by $H_{\text{reg-ch}}$. The unitary transformation \cite{spin2,FT}
\begin{equation}
a_{i}=\sqrt{\frac{2}{N+1}}\sum_{k=1}^{N}\sin\frac{ik\pi}{N+1}b_{k}, \quad i=1,\cdots, N \nonumber
\end{equation}
is implemented to turn $H_{\text{ch}}$ into $H_{\text{ch}}=\sum_{k=1}^{N}\varepsilon_{k}b_{k}^{\dag}b_{k}$, and $\varepsilon_{k}=2g_{C}\cos\left[k\pi/\left(N+1\right)\right]$ is the collective excitation spectrum. In a similar fashion, $H_{\text{reg-ch}}$ is transformed to
\begin{equation}
H_{\text{reg-ch}}=\sum_{k=1}^Nt_{k}\left[a_{L_{n}}^{\dag}b_{k}+\left(-1\right)^{k-1}a_{R_{n}}^{\dag}b_{k}+\text{H.c.}\right],\nonumber
\end{equation}
the coupling between the registers and the $k$th collective mode is parameterized by
\begin{equation}
t_{k}=g_{I}\sqrt{\frac{2}{N+1}}\sin\frac{k\pi}{N+1}.\nonumber
\end{equation}
In a case of odd $N$ chains, we observe a zero-energy collective mode, corresponding to $k=\kappa\equiv\left(N+1\right)/2$, and this mode is in resonance with the registers. By working within the weak-coupling regime, it indicates that $t_{\kappa}\ll|\varepsilon_{\kappa\pm 1}|$ and off-resonant coupling can be neglected. As a consequence, evolution dynamics behaves as an effective model in which only the registers and the zero-energy collective mode are involved. Therefore, $H_{0}$ could be approximated as an effective Hamiltonian of $2n+1$ fermions, as shown in Fig. \ref{f1}(b). Explicitly, this effective Hamiltonian is $H_{\text{eff}}=H_{\text{reg}}+H'_{\text{reg-ch}}$, whereas $H'_{\text{reg-ch}}$ captures an effective coupling between the registers and the channel,
\begin{equation}
H'_{\text{reg-ch}}=t_{\kappa}\left(a_{L_{n}}^{\dag}b_{\kappa}+a_{R_{n}}^{\dag}b_{\kappa}\right)+\text{H.c.},
\end{equation}
and a phase factor of $\left(-1\right)^{\kappa-1}$ has been absorbed into $a_{R_{n}}^{\dag}$. Correspondingly, $\Omega$ is significantly simplified to an $\left(2n+1\right)\times\left(2n+1\right)$ coupling matrix given by
\begin{equation}
\Omega_{\text{eff}}=\left(
\begin{array}{cccccccc}
    0 & g_{1}\\
    \ddots &\ddots & \ddots \\
    &g_{n-1} & 0 & t_{\kappa} \\
    && t_{\kappa} & 0 & t_{\kappa} \\
    &&& t_{\kappa} & 0 & g_{n-1}\\
    &&&& \ddots & \ddots & \ddots \\
    &&&&& g_{1} & 0 \\
\end{array}
\right).\nonumber
\end{equation}
In order to illustrate QST, we set
\begin{equation}
g_{u}=\frac{g_{0}}{2}\sqrt{u\left(2n-u+1\right)},
\end{equation}
where $g_{0}$ is an arbitrary coupling parameter and $u=1,\cdots,n$. Tuning $g_{n}=t_{\kappa}$ implies that $\Omega_{\text{eff}}=g_{0}J_{x}$, and $J_{x}$ is the $x$ component of a pseudo spin $J=n$. By choosing the evolution time $\tau$ such that $g_{0}\tau=\pi$, it allows for
\begin{equation}\label{eq:opevo}
a_{L_{u}}^{\dag}\left(\tau\right)=\left(-1\right)^na_{R_{u}}^{\dag},\ a_{R_{u}}^{\dag}\left(\tau\right)=\left(-1\right)^na^{\dag}_{L_{u}},
\end{equation}
and $b_{\kappa}^{\dag}\left(\tau\right)=\left(-1\right)^nb_{\kappa}^{\dag}$.

The effective fermionic subspace is spanned by
\begin{equation}
\left\{|n_{L_{1}},\cdots, n_{L_{n}},n_{\kappa},n_{R_{n}},\cdots, n_{R_{1}}\rangle\right\},\nonumber
\end{equation}
for $n_{L_{u}},n_{\kappa},n_{R_{u}}=0,1$. In this subspace, the effective evolution for time $\tau$ is
\begin{equation}\label{eq:gevolution}
U_{0}\left(\tau\right)\approx U_{\text{eff}}\left(\tau\right)=\Gamma_{0}\Gamma_{1}\Gamma_{2}
\left(\text{SWAP}_{L,R}\otimes I_{\kappa}\right).
\end{equation}
Here, $U_{0}\left(t\right)=\text{exp}\left(-iH_{0}t\right)$, $U_{\text{eff}}\left(t\right)=\text{exp}\left(-iH_{\text{eff}}t\right)$, and SWAP$_{L,R}$ is a swap operation between the two end registers,
\begin{equation}
\text{SWAP}_{L,R}=\prod_{u=1}^n\text{SWAP}_{L_{u},R_{u}},\nonumber
\end{equation}
and SWAP$_{L_{u},R_{u}}$ swaps the quantum state of the two fermions $L_{u}$ and $R_{u}$. $I_{\kappa}$ is an identity operation acting on the $\kappa$th collective mode of the channel. The additional phase factor of $\Gamma_{0}$ could be solved as
\begin{equation}
\Gamma_{0}=(-1)^{nn_{\kappa}}\prod_{v=1}^{n}(-1)^{n\left(n_{L_{v}}+n_{R_{v}}\right)},\nonumber
\end{equation}
which is independent of fermioninc anticommutaion relations. Meanwhile, such relations give arise to phase factors $\Gamma_{1}=\prod_{x=2}^{n+1}\Gamma_{1}\left(x\right)$ and $\Gamma_{2}=\prod_{x=2}^{n+1}\Gamma_{2}\left(x\right)$ with
\begin{multline}
\Gamma_{1}\left(x\right)=\left[\prod_{v=x}^{n}\left(-1\right)^{n_{L_{x-1}}n_{L_{v}}}\right]\times\\
\left[\prod_{v=1}^{n}\left(-1\right)^{n_{L_{x-1}}n_{R_{v}}}\right]\left(-1\right)^{n_{L_{x-1}}n_{\kappa}},\nonumber
\end{multline}
and
\begin{equation}
\Gamma_{2}\left(x\right)=\left[\prod_{v=x}^{n}\left(-1\right)^{n_{R_{x-1}}n_{R_{v}}}\right]
\left(-1\right)^{n_{R_{x-1}}n_{\kappa}},\nonumber
\end{equation}
respectively. Upon ensuring the weak-coupling, $g_{I}/g_{C}\ll1$, unitary evolution enables a mirror inversion of quantum states of the registers with respect to the center of the channel. However, in addition to the mirror inversion, the registers are entangled with the $\kappa$th collective mode of the channel through the three phase factors. When the channel is in a mixed state, QST will experience the leakage of quantum information into the noisy channel due to this entanglement, thus, QST may be readily inaccessible. Besides the channel noise, QST also suffers from the environment-induced decoherence. To overcome the above drawbacks and use the system in better details, we consider a robust QST protocol by employing a combination of the weak-coupling and the DFSs in the following section.

\section{Decoherence-free subspace construction and Robust Quantum state transfer}
\label{se:section3}
A paradigm has been provided by the use of DFSs as being one of the most important avenues to protect quantum information from decoherence.  Without loss of generality, the effects of decoherence on a system are featured by the following interaction model \cite{dfs_ns_dd}
\begin{equation}
\mathcal{H}_{\text{sys-env}}=\sum_{\alpha}S_{\alpha}\otimes E_{\alpha},\nonumber
\end{equation}
where $S_{\alpha}$ is a pure-system operator, and $E_{\alpha}$ represents an arbitrary pure-environment operator. The environmental Hamiltonian is assumed to be $\mathcal{H}_{\text{env}}$. In a DFS spanned by $\left\{|d_{\gamma}\rangle\right\}$, it requires
\begin{equation}\label{eq:DFS}
S_{\alpha}|d_{\gamma}\rangle=s_{\alpha}|d_{\gamma}\rangle,\quad \text{and} \quad \mathcal{H}_{\text{sys}}|d_{i}\rangle\in \text{DFS},
\end{equation}
which entail that $|d_{\gamma}\rangle$ are eigenvectors of $\mathcal{H}_{\text{sys-env}}$, and the DFS is invariant under the free evolution of the system. The two requirements exhibit that a complete evolution $\mathcal{U}\left(t\right)$ is
\begin{equation}
\mathcal{U}\left(t\right)=\mathcal{U}_{\text{sys}}\left(t\right)\otimes \mathcal{U}_{\text{env}}\left(t\right),\nonumber
\end{equation}
which can be factorized into two parts, the free evolution of the system $\mathcal{U}_{\text{sys}}\left(t\right)=\text{exp}\left(-i\mathcal{H}_{\text{sys}}t\right)$, and the action only on the environment $\mathcal{U}_{\text{env}}\left(t\right)=\text{exp}\left[-i\left(\mathcal{H}_{\text{env}}+\sum_{\alpha}s_{\alpha}E_{\alpha}\right)t\right]$.
Suppose that the initial state of the system $\rho_{\text{sys}}\left(0\right)$ is embedded in the DFS. By tracing out the environmental variables, we arrive at
\begin{equation}
\rho_{\text{sys}}\left(t\right)=\mathcal{U}_{\text{sys}}\left(t\right)\rho_{\text{sys}}\left(0\right)\mathcal{U}_{\text{sys}}^{\dag}\left(t\right).
\end{equation}
As desired, external noise cannot evolve into the DFS, which is decoupled from its surrounding environment to form an isolated subspace.

In particular, we consider a collective dephasing model as being a major source of decoherence in quantum open system. The associated interaction is reduced to $\mathcal{H}_{\text{sys-env}}=S_{z}\otimes E_{z}$ with a collective spin operator
\begin{equation}
S_{z}=\sum_{k}\sigma_{k}^{z},
\end{equation}
which represents the total spin $z$ projection of the system. For $n$ qubits and $M_{\uparrow}$ spins pointing up, eigenvalues of $S_{z}$ are $s_{z}=2M_{\uparrow}-n$ and $M_{\uparrow}=0,\cdots,n$. As required by Eq. (\ref{eq:DFS}), the DFS is the eigenspace of each $s_{z}$, and the dimension is
\begin{equation}
\text{D}_{n}^{M_{\uparrow}}=\frac{n!}{\left(M_{\uparrow}\right)!\left(n-M_{\uparrow}\right)!}.\nonumber
\end{equation}
Because of conservation of total spin $z$ projection, $M_{\uparrow}$ is a constant in each eigenstate for a given DFS. Upon fermionizing, each eigenstate contains $M_{\uparrow}$ creation operators
\begin{equation}
|d_{\gamma}\rangle=\left(a_{\gamma,1}^{\dag}\cdots a_{\gamma,M_{\uparrow}}^{\dag}\right) |0\rangle, \quad \gamma=1,\cdots,D_{n}^{M_{\uparrow}},\nonumber
\end{equation}
where $a_{\gamma,\delta}^{\dag}$ are the creation operators acting on the $\delta$th spin-up in the $\gamma$th eigenstate. In a special case of $M_{\uparrow}=1$, the DFS is spanned by $\left\{|d_{\gamma}\rangle=a_{\gamma,\gamma}^{\dag}|0\rangle\right\}$ and $\gamma$ runs from $1$ to $n$.

Let us now describe our physical model. The initial states of two end registers are encoded in their DFSs,
\begin{equation}
|\varphi\rangle_{L}=\sum_{\gamma=1}^{\text{D}_{n}^{M_{\uparrow}}}\mu_{\gamma}|d_{\gamma,L}\rangle, \quad \text{and} \quad |\phi\rangle_{R}=\sum_{\gamma=1}^{\text{D}_{n}^{M_{\uparrow}}}\nu_{\gamma}|d_{\gamma,R}\rangle, \nonumber
\end{equation}
respectively, where $|d_{\gamma,L/R}\rangle$ is the $\gamma$th basis vector spanning the DFS of the left/right register, while the environment and the channel are in $|\Phi\rangle_{C}$ and $|\Phi\rangle_{E}$. The complete system, including the registers, the intermediate channel and the environment, is in a product state $|\Psi\rangle_{\text{ini}}=|\Phi\rangle_{0}\otimes|\Phi\rangle_{E}$ with $|\Phi\rangle_{0}=|\varphi\rangle_{L}\otimes|\Phi\rangle_{C} \otimes |\phi\rangle_{R}$. Taking into account a collective dephasing, the registers and the channel are exposed to a Hamiltonian
\begin{equation}
H_{\text{int}}=\left[\sum_{i=1}^{N}\sigma_{i}^{z}+\sum_{i=1}^{n}\left(\sigma_{L_{i}}^{z}+\sigma_{R_{i}}^{z}\right)\right]\otimes E_{z}.
\end{equation}
The complete dynamics is governed by a full Hamiltonian $H_{\text{full}}=H_{0}+\mathcal{H}_{\text{env}}+H_{\text{int}}$. Due to $\left[H_{0},H_{\text{int}}\right]=0$, the complete evolution operator $U_{\text{full}}\left(t\right)=\text{exp}\left(-iH_{\text{full}}t\right)$ can be decomposed into two independent processes, the free transport of quantum state in the intermediate channel, $U_{0}\left(t\right)=\text{exp}\left(-iH_{0}t\right)$, and the effects of decoherence, $U_{\text{dep}}\left(t\right)=\text{exp}\left[{-i\left(\mathcal{H}_{\text{env}}+H_{\text{int}}\right)t}\right]$.
By restricting to even $n$ and the weak-coupling regime, $g_{I}/g_{C}\ll1$, it yields $U_{0}\left(\tau\right)|\varphi\rangle_{L}\otimes|\Phi\rangle_{C}\otimes|\phi\rangle_{R}=\left(-1\right)^{M_{\uparrow}}|\phi\rangle_{L}\otimes\left[U_{\text{ch}}\left(\tau\right)|\Phi\rangle_{C}\right]\otimes|\varphi\rangle_{R}$ for a given evolution time $\tau$, implying that
\begin{equation}
U_{0}\left(\tau\right)\approx \left(-1\right)^{M_{\uparrow}}\text{SWAP}_{L,R}\otimes U_{\text{ch}}\left(\tau\right),
\end{equation}
where $U_{\text{ch}}\left(\tau\right)=\text{exp}\left(-iH_{\text{ch}}\tau\right)$. Furthermore, a straightforward calculation gives $U_{\text{dep}}\left(\tau\right)=I_{L}\otimes I_{R}\otimes U'_{\text{dep}}\left(\tau\right)$, and
\begin{equation}
U'_{\text{dep}}\left(\tau\right)=\text{exp}\left\{-i\left[\mathcal{H}_{\text{env}}+\left(\sum_{i=1}^{N}\sigma_{i}^{z}+2s_{z}\right)E_{z}\right]\tau\right\}.\nonumber
\end{equation}
In combination, the complete evolution is
\begin{equation}\label{eq:effevolution}
U_{\text{full}}\left(\tau\right)\approx\text{SWAP}_{L,R} \otimes U_{\text{ch}}\left(\tau\right)U'_{\text{dep}},
\end{equation}
up to an additional global phase. Thus, the finial state becomes
\begin{equation}
|\Psi\rangle_{\text{fin}}=\left(|\phi\rangle_{L}\otimes|\varphi\rangle_{R}\right)\left[U'_{\text{dep}}\left(\tau\right)U_{\text{ch}}\left(\tau\right)|\Phi\rangle_{C}\otimes|\Phi\rangle_{E}\right].\nonumber
\end{equation}
Although our state transfer method is probably fragile to small errors in the intraregister couplings owing to its dependence on the swapping mechanism \cite{spin2,smallerror}, Eq. (\ref{eq:effevolution}) demonstrates that a robust QST against both the channel noise and the environmental decoherence is implemented.

\section{two-qubit registers and Fidelity calculation}
\label{se:section4}
In this section, we take $n=2$ as an example to encode one logical qubit, and then simulate numerically the average fidelity to confirm the efficiency of the method. The average fidelity over the Bloch sphere is defined as \cite{fide1,fide2}
\begin{equation}
F_{\text{avg}}=\frac{1}{2}+\frac{1}{12}\sum_{i=x,y,z}\text{Tr}\left[U\sigma^{i}U^{\dag}\mathcal{M}\left(\sigma^{i}\right)\right],\nonumber
\end{equation}
where $U$ is a unitary transformation and $\mathcal{M}$ is a general trace-preserving operation. As $\mathcal{M}$ implements $U$ perfectly, we observe that $F_{\text{avg}}=1$.
For two physical qubits interacting collectively with a dephasing environment, there exists a two-dimensional DFS
\begin{equation}
\mathcal{S}^{\text{DFS}}=\text{Span}\left\{|\downarrow\uparrow\rangle,|\uparrow\downarrow\rangle\right\},
\end{equation}
and we encode a logical qubit in this subspace as $|0\rangle_{\text{logic}}=|\downarrow\uparrow\rangle$ and $|1\rangle_{\text{logic}}=|\uparrow\downarrow\rangle$. Initially, the physical qubits $L_{1}$ and $L_{2}$ are prepared in $|\varphi\rangle_{L_{1}}=\beta_{0}|\downarrow\rangle+\beta_{1}|\uparrow\rangle$ and $|\downarrow\rangle$, respectively, and then quantum information is encoded into the DFS through a Controlled-NOT (CNOT) operation C$_{L}$ on $L_{1}$ (the control qubit) and $L_{2}$ (the target qubit) to realize an arbitrary logical state $|\varphi\rangle_{L}=\beta_{0}|0\rangle_{\text{logic}}+\beta_{1}|1\rangle_{\text{logic}}$, while the combined state of the channel and the right register is a mixed state $\rho_{\text{ch},R}$. Having been succeed in transferring, quantum information is likewise decoded from the DFS through a CNOT operation C$_{R}$ on $R_{1}$ (the control qubit) and $R_{2}$ (the target qubit), and it enables $|\varphi\rangle_{R_{1}}=\beta_{0}|\downarrow\rangle+\beta_{1}|\uparrow\rangle$. Thus, the average fidelity becomes \cite{weakspin2}
\begin{equation}
F_{\text{DFS}}=\frac{1}{2}+\frac{1}{12}\sum_{i=x,y,z}\text{Tr}\left[\sigma_{R_{1}}^{i}\left(t\right)\left(\sigma_{L_{1}}^{i}\otimes|\downarrow\rangle\langle\downarrow|\otimes \rho_{\text{ch},R}\right)\right],\nonumber
\end{equation}
where $\sigma_{R_{1}}^{i}\left(t\right)=\mathcal{M}^{\dag}\sigma_{R_{1}}^{i}\mathcal{M}$, and $\mathcal{M}=\text{C}_{R}U_{0}\left(t\right)\text{C}_{L}$. Based on
$\text{C}_{R}\sigma_{R_{1}}^{x}\text{C}_{R}=\sigma_{R_{1}}^{x}\sigma_{R_{2}}^{x}$, $\text{C}_{L}\sigma_{L_{1}}^{+}\text{C}_{L}=\sigma_{L_{1}}^{+}\sigma_{L_{2}}^{x}$,
$\text{C}_{L}\sigma_{L_{2}}^{+}\text{C}_{L}
=\frac{1}{2}\left(\sigma_{L_{2}}^{x}+i\sigma_{L_{1}}^{z}\sigma_{L_{2}}^{y}\right)$,
we obtain
\begin{multline}
\text{Tr}\left[\sigma_{R_{1}}^{x}\left(t\right)\left(\sigma_{L_{1}}^{x}\otimes|\downarrow\rangle\langle\downarrow|\otimes \rho_{\text{ch},R}\right)\right] \\=2\text{Re}\left(\Delta_{R_{1},L_{1}}\Delta_{R_{2},L_{2}}^{*}+\Delta_{R_{1},L_{2}}\Delta_{R_{2},L_{1}}^{*}\right).
\end{multline}
Similarity,
\begin{multline}
\text{Tr}\left[\sigma_{R_{1}}^{y}\left(t\right)\left(\sigma_{L_{1}}^{y}\otimes|\downarrow\rangle\langle\downarrow|\otimes \rho_{\text{ch},R}\right)\right] \\=2\text{Re}\left(\Delta_{R_{1},L_{1}}\Delta_{R_{2},L_{2}}^{*}-\Delta_{R_{1},L_{2}}\Delta_{R_{2},L_{1}}^{*}\right).
\end{multline}
By harnessing $\text{C}_{L}\sigma_{L_{1}}^{z}\text{C}_{L}=\sigma_{L_{1}}^{z}$ and $\text{C}_{L}\sigma_{L_{2}}^{z}\text{C}_{L}=\sigma_{L_{1}}^{z}\sigma_{L_{2}}^{z}$, we find
\begin{multline}
\text{Tr}\left[\sigma_{R_{1}}^{z}\left(t\right)\left(\sigma_{L_{1}}^{z}\otimes|\downarrow\rangle\langle\downarrow|\otimes \rho_{C,R}\right)\right] \\=2\left(|\Delta_{R_{1},L_{1}}|^{2}-|\Delta_{R_{1},L_{2}}|^{2}\right).
\end{multline}
Combining the three contributions, we rewrite the fidelity as
\begin{multline}
F_{\text{DFS}}=\frac{1}{2}+\frac{1}{6}\Big[2\text{Re}\left(\Delta_{R_{1},L_{1}}^{*}\Delta_{R_{2},L_{2}}\right)\\
+|\Delta_{R_{1},L_{1}}|^{2}-|\Delta_{R_{1},L_{2}}|^{2}\Big].
\end{multline}
\begin{figure}[!ht]
\begin{center}
\includegraphics[width=8.0cm,angle=0]{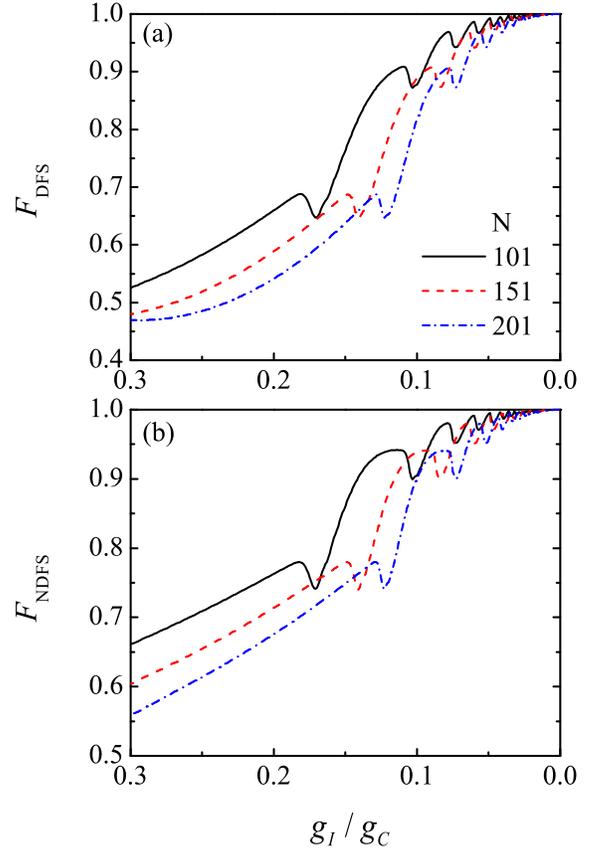}
\caption{(Color online) The average fidelity as a function of $g_{I}/g_{C}$ with either $N=101$ for solid black lines, $N=151$ for dashed red lines, or $N=201$ for dotted-dashed blue lines, in a DFS spanned by $\left\{|\downarrow\uparrow\rangle,|\uparrow\downarrow\rangle\right\}$ (a) and a non-DFS spanned by $\left\{|\downarrow\downarrow\rangle,|\uparrow\uparrow\rangle\right\}$. Here, the evolution time $t=\tau$.}\label{f2}
\end{center}
\end{figure}
The weak-coupling limit results in
\begin{equation}
\Delta_{R_{1},L_{1}}=\frac{1}{8}\left[3-4\cos\left(g_{0}t\right)+\cos\left(2g_{0}t\right)\right],\nonumber
\end{equation}
\begin{equation}
\Delta_{R_{2},L_{2}}=\frac{1}{2}\left[-\cos\left(g_{0}t\right)+\cos\left(2g_{0}t\right)\right],\nonumber
\end{equation}
and
\begin{equation}
\Delta_{R_{1},L_{2}}=\frac{i}{4}\left[2\sin\left(g_{0}t\right)-\sin\left(2g_{0}t\right)\right].\nonumber
\end{equation}
If we set $\Delta_{R_{1},L_{1}}=\Delta_{R_{1},L_{1}}=1$, and $\Delta_{R_{1},L_{2}}=0$ at a given evolution time $\tau$, it leads to $F_{\text{DFS}}=1$, yielding a perfect QST between two end logical qubits. As shown in Fig. \ref{f2}(a), we perform numerics, and the average fidelity varies as a function of $g_{I}/g_{C}$. While encoding in the DFS prevents the leakage of quantum information into the environment, the infidelity results from the off-resonant coupling between the registers and the nonzero-energy collective modes of the channel. Decreasing $g_{I}/g_{C}$ can suppress such irreversible process, and as $g_{I}/g_{C}\approx0$, a robust perfect QST against both the channel noise and the environmental decoherence is achieved.

It is worth noting that a non-DFS subspace spanned by
\begin{equation}
\mathcal{S}^{\text{NDFS}}=\left\{|\downarrow\downarrow\rangle,|\uparrow\uparrow\rangle\right\}
\end{equation}
can also eliminate the unwanted entanglement between the registers and the channel, and result in a robust QST only against the channel noise. In $\mathcal{S}^{\text{NDFS}}$, the time evolution operator is
\begin{equation}
U_{0}\left(\tau\right)\approx \sigma_{\text{logic},R}^{z}\sigma_{\text{logic},L}^{z}\text{SWAP}_{L,R}\otimes I_{C},
\end{equation}
and $\sigma_{\text{logic},L/R}^{z}$ is a logical $z$ operation on the DFS of the left/right register,
\begin{equation}
\sigma_{\text{logic},L/R}^{z}=\left(\sigma_{L_{1}/R_{1}}^{z}+\sigma_{L_{2}/R_{2}}^{z}\right)/2.\nonumber
\end{equation}
The average fidelity is described as
\begin{multline}
F_{\text{NDFS}}=\frac{1}{2}+\frac{1}{6}[2\text{Re}(\Delta_{R_{1},L_{1}}\Delta_{R_{2},L_{2}}\\
-\Delta_{R_{1},L_{2}}\Delta_{R_{2},L_{1}})+|\Delta_{R_{1},L_{1}}|^{2}+|\Delta_{R_{1},L_{2}}|^{2}],
\end{multline}
with $\Omega_{R_{2},L_{1}}=\Omega_{R_{1},L_{2}}$. Obviously, $g_{I}/g_{C}\ll1$ reveals $F_{\text{NDFS}}=1$. Numerical simulations are performed as plotted in Fig. \ref{f2}(b), and the infidelity results from the leakage of information into the off-resonant modes of the channel, in direct analog to the DFS. However, the non-DFS subspace $\mathcal{S}^{\text{NDFS}}$ cannot provide quantum information protection from the environmental decoherence. In addition, all of the remaining four subspaces spanned by $\left\{|\downarrow\downarrow\rangle,|\downarrow\uparrow\rangle\right\}$, $\left\{|\downarrow\downarrow\rangle,|\uparrow\downarrow\rangle\right\}$, $\left\{|\uparrow\uparrow\rangle,|\downarrow\uparrow\rangle\right\}$, and
$\left\{|\uparrow\uparrow\rangle,|\uparrow\downarrow\rangle\right\}$ are sensitive to either the channel noise or the environmental decoherence, and incapable of encoding information.

\section{Summary}
\label{se:section5}
In this paper, we have investigated a robust QST in a random unpolarized coupled-spin chain of arbitrary length under decoherence. Utilizing a combination of maintaining the weak register-channel coupling and encoding quantum information in the DFSs, the present protocol is immune to both the channel noise and the environmental decoherence, in contrast to previous work. The key ingredient is that multi-qubit quantum states can be efficiently transferred via a long-range coherent quantum interaction between two remote registers in the weak register-channel coupling limit, and it predicts intrinsically the ability to construct the noiseless DFSs by means of multiple qubits. A special case where each register consists of two qubits is taken as an example to calculate explicitly its average fidelity expressed in terms of elements of the evolution matrix, and then the robustness is confirmed by using numerical simulations. It is a direct application which provide potentially the protected entanglement distribution for quantum computation. With its robustness and scalability, the protocol can be further applied to quantum information processing.
\section*{Acknowledgement}

This work is supported by the China National Natural Science Foundation Grant Nos. 61205117 and 61471050, Beijing Higher Education Young Elite Teacher Project No. YETP0456.


\end{document}